# Van der Waals interaction affects wrinkle formation in two-dimensional materials

Pablo Ares[1,*], Yi Bo Wang[1], Colin R. Woods[1], James Dougherty[1], Laura Fumagalli[1], Francisco Guinea[2,3,4], Benny Davidovitch[5,†], Kostya S. Novoselov[1,6,7,‡]

[1]Department of Physics & Astronomy and National Graphene Institute, University of Manchester, Manchester, M13 9PL, UK

[2]IMDEA Nanoscience, C/ Faraday 9, Madrid, 28049, Spain

[3]Donostia International Physics Center, Paseo Manuel de Lardizabal 4, 20018, San Sebastián, Spain

[4]Ikerbasque, Basque Foundation for Science, 48009, Bilbao, Spain

[5]Department of Physics, University of Massachusetts Amherst, Amherst MA 01003, USA

[6]Centre for Advanced 2D Materials, National University of Singapore, 117546, Singapore

[7]Chongqing 2D Materials Institute Liangjiang New Area Chongqing, 400714, People's Republic of China

[*]Corresponding author: pableras.ares@gmail.com

[†]Corresponding author: bdavidov@umass.edu

[‡]Corresponding author: kostya@nus.edu.sg

**Nonlinear mechanics of solids is an exciting field that encompasses both beautiful mathematics, such as the emergence of instabilities and the formation of complex patterns, as well as multiple applications. Two-dimensional crystals and van der Waals (vdW) heterostructures allow revisiting this field on the atomic level, allowing much finer control over the parameters and offering atomistic interpretation of experimental observations. In this work, we consider the formation of instabilities consisting of radially-oriented wrinkles around mono- and few-layer "bubbles" in two-dimensional vdW heterostructures. Interestingly, the shape and wavelength of the wrinkles depend not only on the thickness of the two-dimensional crystal forming the bubble, but also on the atomistic structure of the interface between the bubble and the substrate, which can be controlled by their relative orientation. We argue that the periodic nature of these patterns emanates from an energetic balance between the resistance of the top membrane to bending, which favors large wavelength of wrinkles, and the membrane-substrate vdW attraction, which favors small wrinkle amplitude. Employing the classical "Winkler foundation" model of elasticity theory, we show that the number of radial wrinkles conveys a valuable relationship between the bending rigidity of the top membrane and the strength of the vdW interaction. Armed with this relationship, we use our data to demonstrate a nontrivial dependence of the bending rigidity on the number of layers in the top membrane, which shows two different regimes driven by slippage between the layers, and a high sensitivity of the vdW force to the alignment between the substrate and the membrane.**

The extraordinary mechanical properties of two-dimensional (2D) materials have attracted considerable interest from the physics and material science communities[1-4]. Such materials present a very versatile playground allowing realization of multiple modelling objects, such as the thinnest possible membranes with miniscule bending rigidity or very stiff plates, all with very reproducible parameters. Furthermore, both the electronic and mechanical properties of such crystals can be fine-tuned by assembling them into van der Waals (vdW) heterostructures[5-7], which allow realization of yet even broader range of materials parameters and boundary conditions. Of special interest are the properties of free-standing mono- and few-layer 2D crystals: membranes and bubbles[8-10]. The elastic response of such systems is highly nontrivial, being governed by the vdW membrane-substrate's attraction, the bending rigidity of 2D solid membranes and their Young modulus. Furthermore, atomic thinness of the membranes makes it possible to switch between different regimes – from where the bending rigidity is much smaller than the Young modulus (monolayer membrane), to where they are comparable (in the case of few-layer membranes). A common outcome of this interplay is the instability of the neutrally planar state and the consequent emergence of complex wrinkle patterns when the membrane experiences minute levels of compression[11-13]. Understanding the conditions under which such instabilities emerge and spread throughout the system is of utmost importance both fundamentally and for the functionality of various vdW heterostructure devices. Specifically, the formation of wrinkling patterns is usually associated with a nontrivial strain distribution[14,15], which, for piezoelectric 2D crystals such as $MoS_2$ and hBN[16-18], leads to a complex distribution of the electric field, and can be used in respective devices. Furthermore, a careful analysis of the emerging patterns can be harnessed for metrological purposes – extracting valuable information on the material parameters that characterize the elastic modulii of 2D membranes and their vdW interaction with the substrate.

In this work, we study patterns of radially-oriented wrinkles induced by bubbles in vdW heterostructures that consist of hexagonal boron nitride (hBN) or molybdenum disulfide ($MoS_2$) layers on top of another 2D material. We show that the periodicity of the emerging patterns reflects a balance between the membranal bending rigidity ($B$) and the steepness of the vdW potential well, namely, its second derivative ($V_{vdW}''$), evaluated at the equilibrium value of the membrane-substrate distance[19]. Although this type of balance between bending rigidity and stiffness of an "effective substrate" has been known to govern wrinkle patterns in thin polymer sheets[14,20-24], the current report provides a non-invasive probe for the vdW interaction between atomically thin membranes and atomically flat substrates. Furthermore, the relative crystallographic orientation between the crystalline, atomically thin membrane and the substrate can be controlled precisely, giving us an extra knob to change the vdW interaction. Our results demonstrate that classical nonlinear mechanics of solids, which describes strain-induced elastic instabilities, can be applied (with some modifications) to the description of pattern formation in complex vdW heterostructures and provide a novel metrological tool for characterizing the interplay between the vdW attraction and the bending modulus of monolayers and multi-layer composites.

We produced high-quality vdW heterostructures with atomically flat monocrystals of hBN on top of graphene (both in incommensurate and commensurate configurations) and $MoS_2$ on top of hBN using the standard dry transfer technique[25]. In brief, we mechanically exfoliate and identify our top 2D material on a polymer film consisting of two sacrificial layers, one of polymethylglutarimide (PMGI) and the other of poly(methyl methacrylate) (PMMA). We develop the PMGI layer from beneath the PMMA layer to create a free-standing membrane that we could easily manipulate with the crystal on top. Then we inverted the membrane and we positioned it above our substrate 2D material, which is resting on a SiO2/Si substrate, using a set of micromanipulation stages with spatial accuracy better than 5 μm and alignment precision ~0.5° to control the twist angle. For the commensurate configuration, we aligned the crystal lattices by choosing straight edges of both top and substrate 2D crystals, which indicate the principal crystallographic directions, and brought them into contact making their edges parallel. For the incommensurate configuration, we rotated the top crystal by ~15° with respect to the 2D substrate edges before bringing them into contact. We then removed the PMMA by simply peeling back the membrane. In such structures, bubbles of trapped substances are spontaneously formed between the substrate and the membrane, such that the membrane portion on top of the bubble is completely delaminated from the substrate. Bubbles appear when 2D crystals with high enough adhesion are brought together due to a self-cleansing mechanism[7,26], which pushes trapped contaminants to gather into bubbles[9,27]. We imaged our samples with atomic force microscopy (AFM)[28,29] using a dynamic mode[30,31], so the images were taken in a gentle non-invasive way ensuring that the sample was not damaged or deformed. We found that the typical radii $R$ of bubbles is in the range 50-300 nm, and their heights $H$ are 10-30 nm, however the aspect ratio *H/R* is constant for each given membrane thickness. This constant ratio was shown to be $C_0(\frac{\Gamma}{Y})^{1/4}$, where $C_0$ is a numerical constant, $\Gamma$ is the membrane-substrate surface energy (*i.e.* magnitude of vdW attraction), and $Y \gg \Gamma$ is the 2D membranal stretching modulus[9].

We start our analysis with the incommensurate configuration. Figure 1 shows the development of a "corona" of wrinkles (azimuthal undulations of the shape at the laminated portion around the bubble) for membranes of different thicknesses.

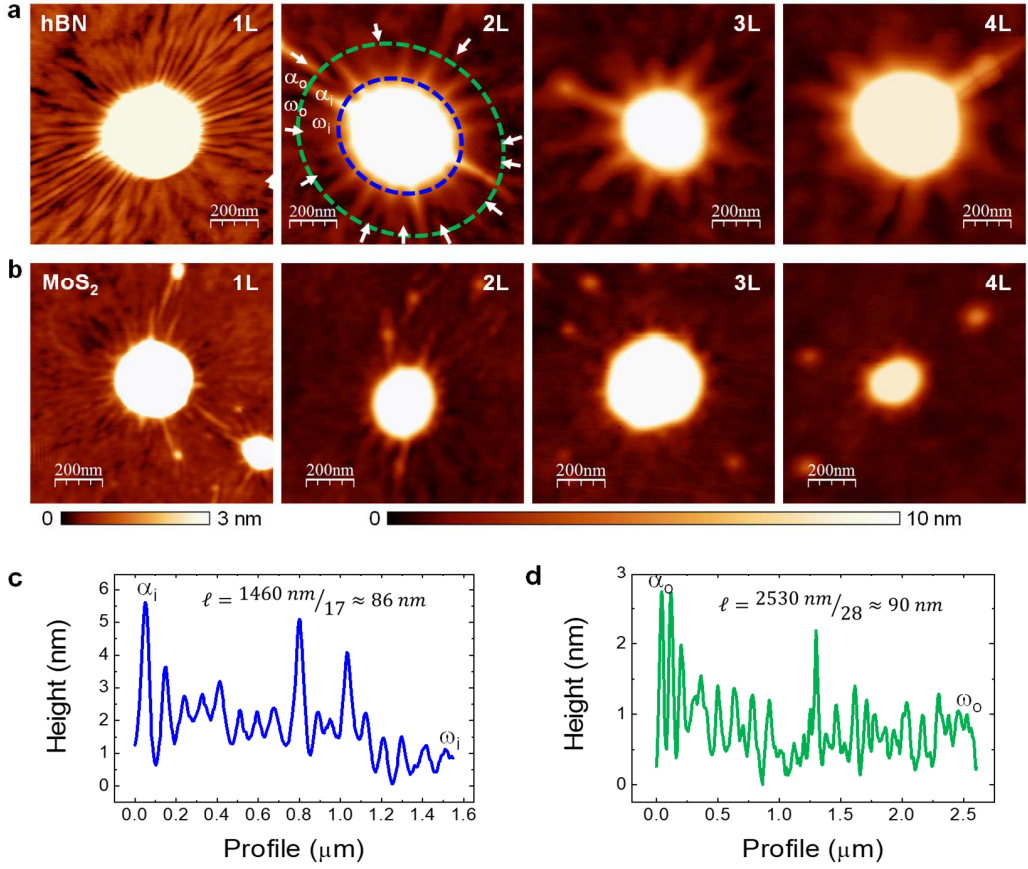

**Fig. 1. Topographic images of wrinkles around bubbles in hBN and MoS$_2$ of different thickness**. **a**, Bubbles with wrinkles in hBN membranes of different thicknesses (monolayer - 1L, bilayer - 2L, trilayer - 3L and four-layer - 4L), incommensurately stacked on a graphite substrate. The arrows in the 2L hBN bubble point to wrinkles that nucleate away from the bubble's edge, such that the condition $m(r) \propto r$ (implied by a spatially-uniform wavelength) is fulfilled. The letters α and ω mark wrinkles for identification in **c** and **d**. **b**, Bubbles with wrinkles in MoS$_2$ membranes of different thicknesses (monolayer - 1L, bilayer - 2L, trilayer - 3L and four-layer - 4L), incommensurately stacked on a thick hBN substrate. Height scales have been adjusted to enhance the wrinkles visibility (3 nm for 1L and 10 nm for 2L, 3L and 4L). **c**, Height profile along the inner (blue) dashed line in **a**. **d**, Height profile along the outer (green) dashed line in **a**.

From profiles around the topographical images of the wrinkles around the bubbles we can determine the wavelength $\ell$ of the wrinkles as $\ell(r) = 2\pi r / m(r)$, where $m(r)$ is the number of azimuthal undulations at a distance $r$. Figure 1c and 1d show examples of the wavelength determination for the case of bilayer hBN for two different radii, yielding similar values (see Supplementary Information Fig. 1 for information on how the wrinkles were counted).

The mere existence of radial wrinkles is attributed to the pressure in the bubble. Rather than a uniform, isotropic surface tension (as would be for a gas-filled vesicle), such a pressure gives rise to an anisotropic stress field in the membrane, whose radial component is tensile, $\sigma_{rr} \sim \Gamma^{1/2} Y^{1/2} \gg \Gamma$, provided that $Y \gg \Gamma$. The radial force is pulling inward the portion of the membrane that remains attached to the substrate ($r > R$), and consequently causing there azimuthal (hoop) compression, $\sigma_{\theta\theta} < 0$ [9]. Since they emerge just in order to suppress such a compression, the radial extent of wrinkles corresponds to the region of azimuthal compression, which is known to be proportional to the bubble's radius R, where the proportionality constant is the ratio $\sigma_{rr}(R)/\sigma_\infty$ between the radial tension $\sigma_{rr}(R) \sim \Gamma^{1/2} Y^{1/2}$ that pulls inward at the bubble's edge and the far-field (uniform, isotropic) stress $\sigma_\infty$ in the membrane[24]. This is illustrated in Fig. 2, where we show that the formation of these radially-oriented wrinkles is correlated with the presence of thin layers of trapped molecules in the vicinity of the bubbles, as observed by monitoring the same bubbles just after fabrication in ambient conditions and after leaving them overnight at high humidity (Relative Humidity, RH $\gtrsim$ 90%) (Fig. 2a-d). A plausible reason for the strong effect of humidity, schematically depicted in Fig. 2e-h, is that the intake of water molecules between the top and the bottom layers acts to increase the pressure inside the bubble, thereby enhancing the radial tension in the top membrane and consequently the hoop compression induced by it, promoting the development of radial wrinkles around the bubble.

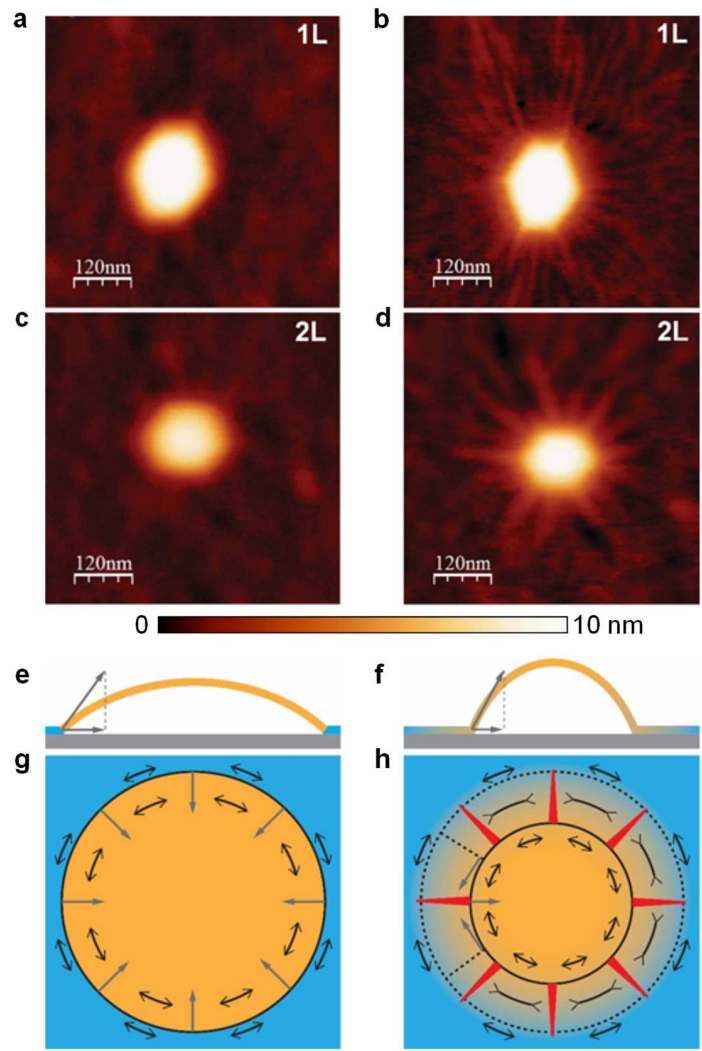

**Fig. 2. Formation of wrinkles. a**, **c**, Bubbles in hBN membranes of different thicknesses (**a** – 1L, **c** – 2L) incommensurately stacked on a graphite substrate just after fabrication. **b**, **d**, Same bubbles as in **a**, **c**, respectively, after leaving them overnight at high humidity, where the appearance of wrinkles is clearly visible. **e**, **g**, Side and top schematic views respectively on a bubble without wrinkles (stress does not spread into the material in contact with the substrate). **f**, **h**, Side and top schematic views respectively on a bubble with wrinkles (stress spreads into the material in contact with the substrate). Shades of yellow represent the level of stress. Single-headed grey arrows represent the elastic forces acting on the perimeter of the bubble. Double-headed black arrows depict components of the stress tensor. The red triangles in **h** represent the wrinkles. Whereas in **g** the azimuthal (hoop) stress is purely tensile, in **h** it is tensile-compressive-tensile.

Throughout the azimuthally-compressed corona of the bubble, the low bending rigidity of the membrane underlies an elastic instability, whereby the membrane seeks to deflect away from the substrate in order to suppress hoop compression and eliminate the large energetic cost it entails. If the vdW attraction were sufficiently weak, one may expect this instability to give rise to a highly non-uniform pattern, where a complete detachment from the substrate in a few narrow angular sectors enables the membrane to retain a full contact with the substrate elsewhere. Such a delamination pattern would be characterized by a small number of narrow, high-amplitude and radially-oriented "folds", separated by perfectly flat, non-undulatory zones, as was assumed in Ref. 11, but in sharp contrast with our observations. Instead, the emergence of a simple periodic pattern, characterized by a single wavelength, suggests another instability mechanism, whereby the vdW attraction is sufficiently strong to resist delamination anywhere but beneath the highly-pressurized bubble, and small-amplitude radial wrinkles emerge to suppress hoop compression in the corona (see schematic Fig. 3a). This phenomenon resembles the effect of a liquid drop placed on top of a floating polymer sheet, where the Laplace pressure inside the drop gives rise to preferentially-radial tension in the sheet, and to the suppression of would-be hoop compression through radial wrinkles[21,24].

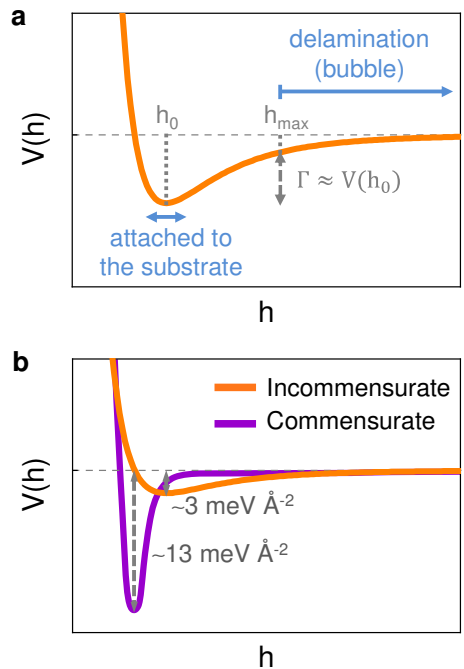

**Fig. 3. Schematic graphs of the vdW potentials. a**, Shallow vdW potential for the case of incommensurate stacking. For substrate-membrane distances above a threshold value ($h_{max}$) the vdW attraction $\Gamma \approx V(h_0)$ is weak and delamination (underlying the formation of bubbles) occurs. In contrast, small-amplitude wrinkles attached to the substrate appear to suppress hoop compression around bubbles, where the vdW attraction is stronger. **b**, Comparison of the vdW potential for the incommensurate and commensurate cases. Our experiments suggest that in the commensurate case the overall magnitude of the vdW attraction, $\Gamma \approx V(h_0)$, increases by ~4, and the steepness of the vdW potential, $K_{eff} \approx V''(h_0)$, increases by ~150.

The typical "wavelength" $\ell$ of these wrinkles depends on the thickness of the membranes (see Fig. 4a). Denoting the membrane-substrate distance by $z(r,\theta)$, the wrinkle pattern can be modeled as:

$$z(r,\theta) = z_0(r) + A(r)\cos[m(r)\theta] \quad (1)$$

Where $A(r)$ and $m(r) = 2\pi r/\ell(r)$ are, respectively, the amplitude and number of azimuthal undulations at a distance $r > R$ from the center of the bubble. From the AFM images we can characterize the shape $z(r,\theta)$, and observe a few noteworthy features:

a. The number of wrinkles decreases sharply with the number of layers N.
b. The wrinkle number $m(r)$ increases with radial distance $r$. Furthermore, a detail inspection indicates that $m(r) \propto r$, such that wrinkles are "nucleated" at various locations[32], and the wavelength $\ell(r) = 2\pi r/m(r)$ is close to a constant value, $\ell(r) = \ell_0$ (see Supplementary Information Fig. 2).
c. In the vicinity of the bubble, $r \geq R$, the non-oscillatory component of the shape, $z_0(r)$, decays exponentially, such that most of the wrinkled zone is nearly flat in the radial direction, and highly curved in the azimuthal direction (Supplementary Information Fig. 3).
d. Notwithstanding their high azimuthal curvature, the amplitude of wrinkles, $A(r)$, is smaller than *1 nm* in the majority of the wrinkled zone (Supplementary Information Fig. 3).
e. For a given type (and number of layers) of the top membrane, the average wavelength $\ell_0$ does not depend on the size (radius or height) of bubbles (Fig. 4a-c).

Focusing our attention on the wavelength of wrinkles, we recall a general relation, known as the "local wavelength law"[14,23]:

$$\ell(r) \approx 2\pi \left(\frac{B_{eff}}{K_{eff}(r)}\right)^{1/4} \quad (2)$$

Here $B_{eff}$ is the effective bending rigidity of the membrane and $K_{eff}(r)$ is an "effective stiffness". Eq. (2) states that the wavelength of wrinkles reflects an energetic balance between the bending rigidity of the membrane, which favors undulations with small curvature (hence large $\ell$) and the restoring forces that favor small amplitude (hence small $\ell$). Physically effective stiffness can originate from multiple contributions, such as attraction to the substrate (vdW interaction), as well as tension- and curvature-induced stiffness (associated with energy cost for deformation of curved membranes). Our analysis (see Supplementary Information for details) indicates that the dominant contribution is the vdW interaction with the substrate $K_{eff} \approx V_{vdW}''(h_0)$. The quantity "effective stiffness" does not have the conventional units of other measurables that are used to characterize stiffness in contact mechanics. Instead, we use the term "effective stiffness" as it is typically employed in the elasticity literature, where it refers to a "spring constant" associated with the resistive force exerted by a unit area of a compliant substrate. Hence the units are $\frac{[Force]/[Length]}{[Area]} = \frac{[Energy]}{[Length]^4}$. In our analysis, this term encapsulates the substrate resistance to the formation of ripples in the layer above it (thereby setting the wavelength through a balance with the bending rigidity of the sheet). In implementing this concept in our study, the relevant energy scale is not the minimum of the substrate-layer attraction energy/area, but rather the penalty for deviations from the minimal value (*i.e.* second derivative of $\frac{Energy}{Area} \to \frac{Energy}{Lentgh^4}$).

Having established the nature of the effective stiffness in Eq. (2), let us address now the bending modulus $B$. Generally, there are two regimes for the bending of multilayer 2D crystals. At low bending curvature (large $\ell$) there is no slippage between the planes and the general continuous mechanics can be applied. For high bending curvature (low $\ell$) when the layers are allowed to slip with respect to each other - the layers behave independently, so the bending rigidity is simply given by the sum of the individual bending rigidities of individual monolayers comprising the membrane. We define an effective bending rigidity, $B_{eff}$, as:

$$B_{eff}(\ell) \approx \begin{cases} \frac{N^3 - N}{12}(\lambda + 2\mu)d^2, & \ell^2 \gg \frac{4\pi^2(N^3 - N)d^2(\lambda + 2\mu)}{12(N-1)\Gamma_s} \\ N \times B, & \ell^2 \ll \frac{4\pi^2(N^3 - N)d^2(\lambda + 2\mu)}{12(N-1)\Gamma_s} \end{cases} \quad (3)$$

where $N$ is the number of layers, $\lambda$ and $\mu$ are in-plane elastic constants (Lamé coefficients), *d* is the distance between layers, *B* is the bending rigidity of a single layer, and $\Gamma_s$ determines the shear mode of multilayered samples. At long wavelengths, the mutual slippage between the layers is suppressed, and the effective bending rigidity scales as the cube of the number of layers, being determined by the in-plane bulk modulus, in agreement with the general theory of elastic slabs in three dimensions [33]. At short wavelengths intra-layer slippage is prominent, such that the mechanical response of each layer is practically independent on other layers, and consequently the bending rigidity equals to the number of layers multiplied by the bending rigidity of a single layer. The crossover wavelength $\ell^*$ is determined by the shear stiffness and takes place at wavelengths and energies that belong to the lowest energy mode with out of plane modulations. The derivation of Eq. (3) is given in the Supplementary Information.

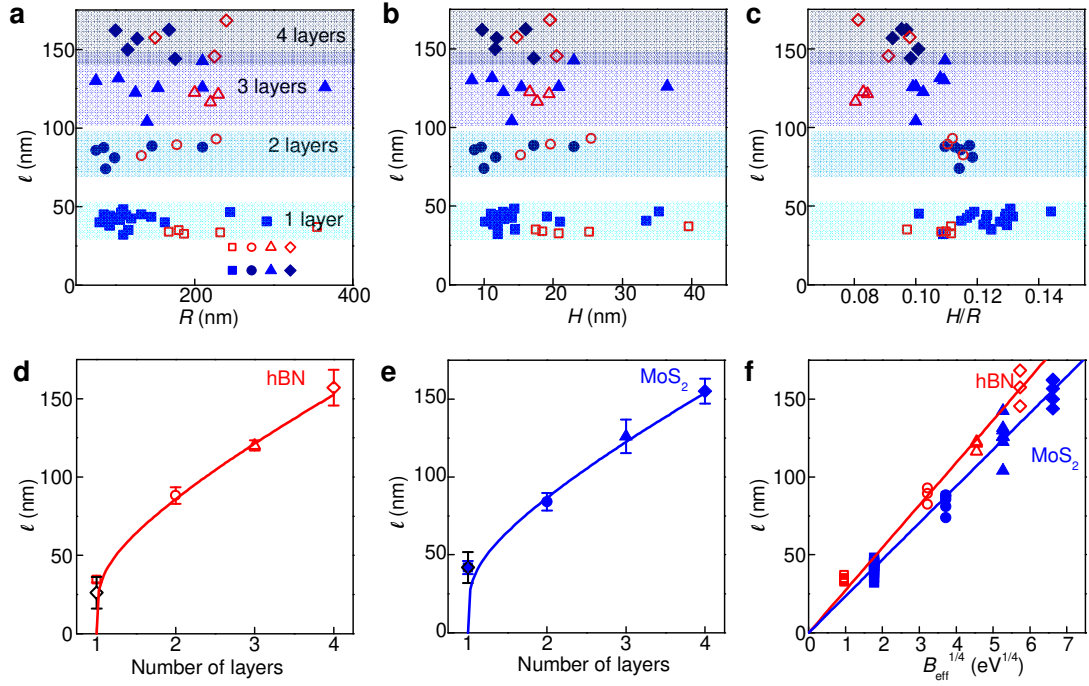

**Fig. 4. Wrinkle's periodicity for hBN and MoS$_2$ bubbles of different thicknesses.** Plots of the average wrinkle wavelength as a function of: **a**, bubble radius, **b**, height and **c**, aspect ratio, for hBN (open red symbols) and MoS$_2$ (solid blue symbols). Here squares are for monolayer membranes, circles – bilayer, triangles – 3-layer, rhombuses – 4-layer. **d**, **e**, Average wavelength as a function of the number of layers (symbols) and fits to Eq. (2) (lines) where the value of $B_{eff}$ is obtained from Eq. (3) for N > 1. Error bars are standard deviation. The open black diamonds correspond to the calculated wavelengths for N = 1, estimated using the effective substrate stiffness values from the line fits, with their error bars coming from propagating the uncertainties in the fits. **f**, Wavelength as a function of $B_{eff}^{1/4}$ (symbols) and fits to Eq. (2) (lines).

Figures 4d and 4e show the wavelength as a function of the number of layers for hBN and MoS$_2$ respectively, and fits of the experimental points to the "local wavelength law" (Eq. (2)), with substrate-dominated stiffness. For these fits, the bending rigidity has been considered as the effective bending rigidity above the crossover length from Eq. (3), thus it scales as the cube of the number of layers, making $\ell \propto N^{3/4}$. Note that Eq. (3) implies that above the crossover $\ell^*$ the wavelengths tend to zero for N = 1 and hence the fits are valid for N > 1 only. We have used the Lamé coefficients $\lambda$ and $\mu$ for hBN and MoS$_2$ given by Refs. 34-37 (see Table S1 in the Supplementary Information for a compilation of the employed elastic constants). From the fits we obtain the effective substrate stiffness for each material, resulting in $K_{eff}$ (hBN) = 2.8×10$^{-7}$ eVÅ$^{-4}$, and $K_{eff}$ (MoS$_2$) = 5.1×10$^{-7}$ eVÅ$^{-4}$. Using these values for the effective substrate stiffness we can obtain the expected wavelengths for the case N = 1, substituting the monolayer bending rigidities for each material into Eq. (2) (black diamonds in Figs. 4d and 4e, compatible with the experimental observed wavelengths). Figure 4f summarizes the results for both materials, where we plot the wavelength as a function of $B_{eff}^{1/4}$ and fit the data to Eq. (2). We find excellent agreement between the theory and the experimental observations.

Let us turn now to commensurate configurations, where the hBN membrane is crystallographically aligned with the underlying graphene membrane. In this case, the behavior is remarkably different. Small rotation angles between the lattices of the two crystals form a periodic structure with different stacking configurations, yielding a moiré pattern[38]. Figure 5 corresponds to a monolayer of hBN aligned on top of a graphene flake presenting bubbles, as in the incommensurate states. It can be observed, however, that in the commensurate case the wrinkles exhibit a substantial deviation from radial orientation, and become constrained to the moiré periodicity originated by the strong interaction between the top and the bottom layers. The wavelength of the wrinkles in this case is $\ell$ ~ 7.8 nm, the same as the moiré periodicity, and they follow preferential 60° directions, presenting abrupt kinks where it is necessary to accommodate the underlying moiré pattern.

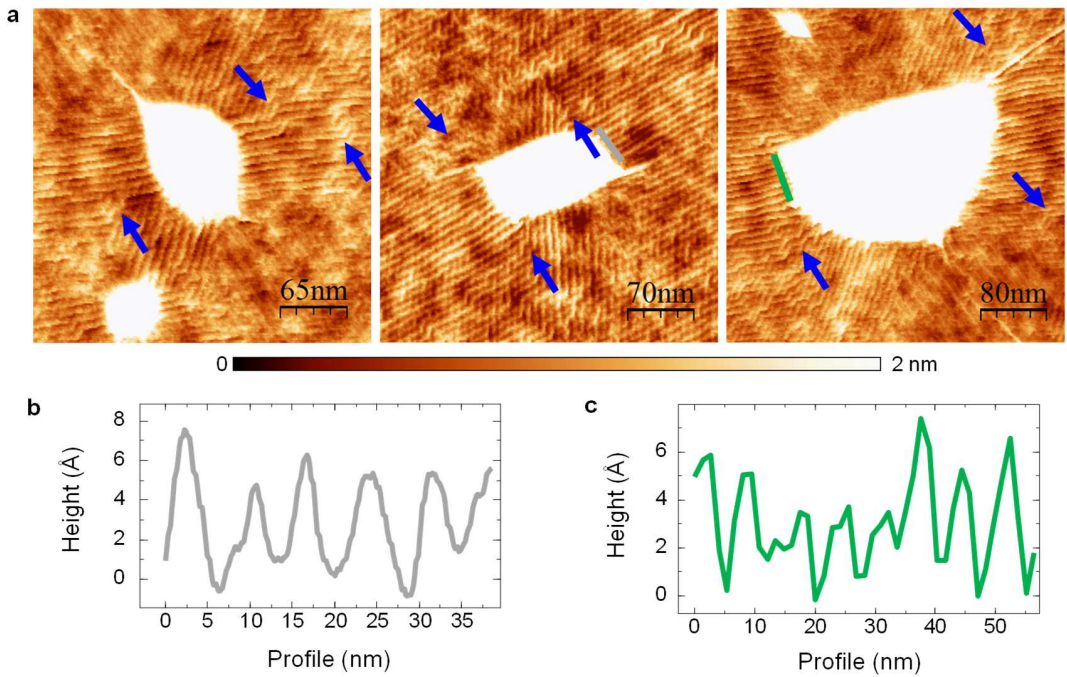

**Fig. 5. Wrinkles in commensurate case. a**, Wrinkles around bubbles in commensurate heterostructures of aligned monolayer hBN on graphene. The arrows point to kinks where the wrinkles change their orientations abruptly, following the underlying moiré pattern. **b**, **c**, Profiles along the lines in the middle (**b**) and right (**c**) panels of **a**. The height of the wrinkles in the commensurate cases is ~3-6 Å.

This higher interaction is directly reflected in the aspect ratio of the bubbles. Whereas the typical *H/R* ratio for the incommensurate state of a monolayer hBN on graphene is 0.11 ± 0.02, for the commensurate state this value increases to *H/R* = 0.16 ± 0.02 (see Supplementary Information Fig. 4). As mentioned in the introduction, the aspect ratio *H/R* is proportional to the ratio $(\frac{\Gamma}{Y})^{1/4}$, where $\Gamma$ is the membrane-substrate surface energy and $Y \gg \Gamma$ is the 2D membranal stretching modulus. In particular, from the measured values of the aspect ratio, *H/R*, and using a Young modulus $Y$ = 289 N m$^{-1}$ = 18 eV Å$^{-2}$ for monolayer hBN [3], we can obtain the van der Waals interaction, $\Gamma$, from $\frac{H}{R} = C_0 \left(\frac{\Gamma}{Y}\right)^{1/4}$ [9], for both incommensurate and commensurate cases, resulting in $\Gamma_{incomm} = 0.003$ eV Å$^{-2}$ and $\Gamma_{comm} = 0.013$ eV Å$^{-2}$ (Fig. 3b). The value of $\Gamma_{incomm}$ is in line with other estimates of the van der Waals energy in similar setups[9]. The membrane-substrate surface energy in the commensurate state is ~4 times higher than for its counterpart in the incommensurate state (see Fig. 3b, which shows schematics of the vdW potential for the commensurate and incommensurate cases, highlighting the pronounced difference in their respective profiles). This stronger interaction is likely to drastically decrease the presence of water molecules (outside the bubble) between the membrane and the substrate, in comparison to the incommensurate state. This enhanced interaction is also reflected in the fact that the wrinkles in the commensurate state present almost atomic scale heights (see Figs. 5b and c). Carrying out a similar analysis as for the incommensurate case, our experimental observations in the commensurate state indicate that the effective stiffness $K_{eff}$, and consequently the vdW interaction with the substrate, is substantially enhanced in the commensurate case. From Eq. (2) we can calculate $K_{eff}$ for this case, resulting in $K_{eff}$ (hBN)$_{comm}$ ~ 4×10$^{-5}$ eVÅ$^{-4}$, which is two orders of magnitude higher than for the incommensurate state (please note the different steepness of the vdW potential in Fig. 3b). This may not be surprising, since the presence of water and other molecules in the incommensurate wrinkles may "soften" the interaction. The higher interaction in the commensurate case, reflected in a significantly higher magnitude of the vdW attraction and the effective stiffness, suggests that commensurate moiré structures of lengths on the order of 10 nm or larger are very stable.

Our experimental observations portray a nontrivial, multi-scale interplay of the vdW substrate-membrane attraction, the high membranal resistance to stretching and its low bending rigidity, driven by the presence of trapped substances between the membrane and the substrate. The primary response, which has been reported previously and is barely affected by bending rigidity, is the aggregation of trapped substances into highly-pressurized bubbles of radius *R*, thereby causing delamination of the membrane from the substrate at $r \ll R$ (where $h(r) > h_{max}$) in Fig. 3, and consequently a large radial tension, $\sigma_{rr} \sim \Gamma^{1/2} Y^{1/2} \gg \Gamma$ in the delaminated portion of the membrane. The primary response, which is evidently governed by the membranal stretching resistance (through the Young's modulus *Y*) and the overall magnitude of the vdW attraction ($\Gamma \approx V(h_0)$), underlies a secondary effect, due to the highly anisotropic stress in the solid membrane, whereby a hoop compression induced by the radial tension in the vicinity of the bubble is suppressed with the aid of radial wrinkles. The consequent suppression of the elastic energy is governed by much lower contributions, associated with the membranal bending modulus *B* and the steepness of the vdW potential, $K_{eff} \approx V''(h_0)$. Thus, measurements of the various features of the deformed membrane can be harnessed to provide a very convenient, indirect metrological probe. In the current study, we employed this system to determine the dependence of the bending rigidity on the number of atomic layers in the top membrane, and to examine the strong effect of commensurability and humidity on the magnitude and steepness of the vdW potential. We expect that future studies will advance further this novel approach for metrology of vdW heterostructures.

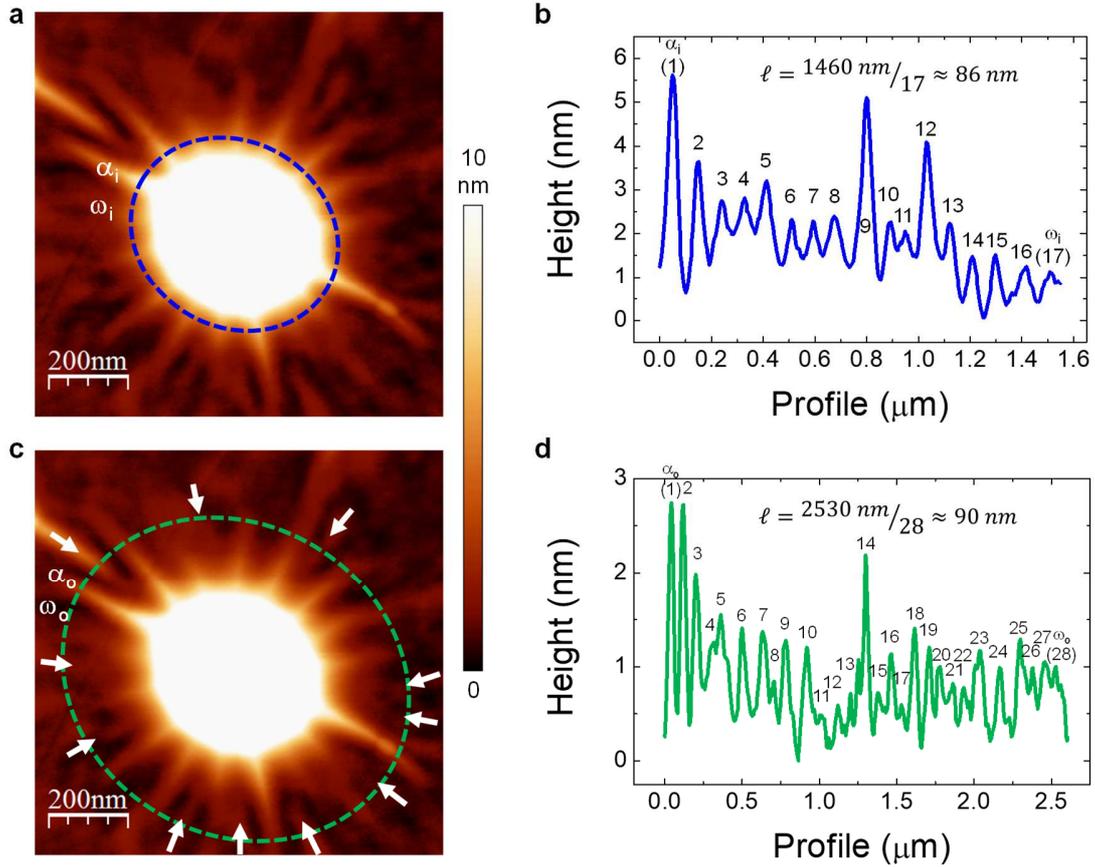

**Fig. S1. Determination of the wrinkle number m(r) for wavelength calculation.** Panels reproduced from Fig.1 in the main text for clarity. **a**, and **c**, Topographic image of wrinkles around a bubble in 2L hBN. Height scale: 10 nm. **b**, and **d**, Height profiles along the dashed lines in **a**, and **c** respectively. The letters α and ω refer to the first and last wrinkles in the profiles. Each wrinkle has been numbered for clarity.

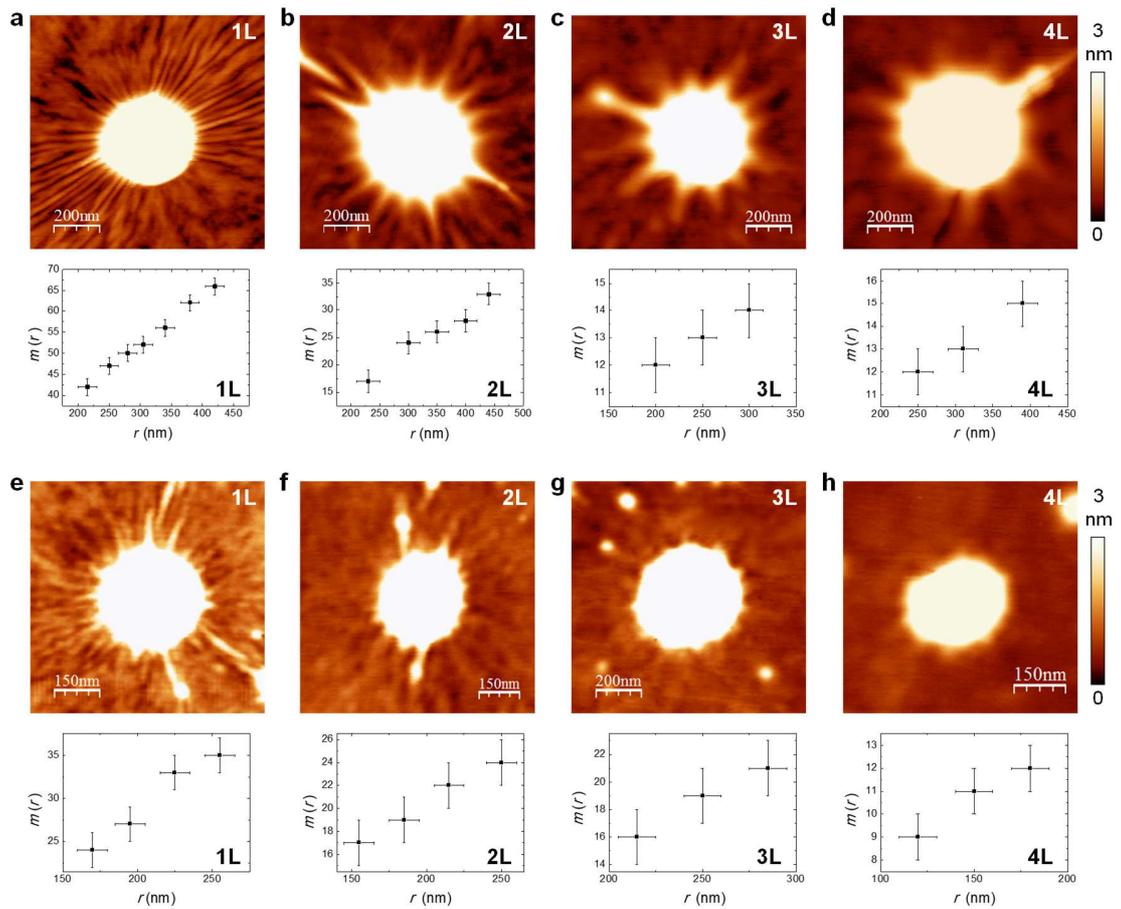

**Fig. S2. Evolution of the wrinkle number *m(r)* with radial distance *r*. a-d**, hBN and **e-h**, MoS2 topographic images (top rows) of bubbles with wrinkles for different layer thicknesses and corresponding wrinkle number *vs.* radial distance (bottom rows). In all the cases, the wrinkle number increases with the radial distance. Height scale: 3 nm.

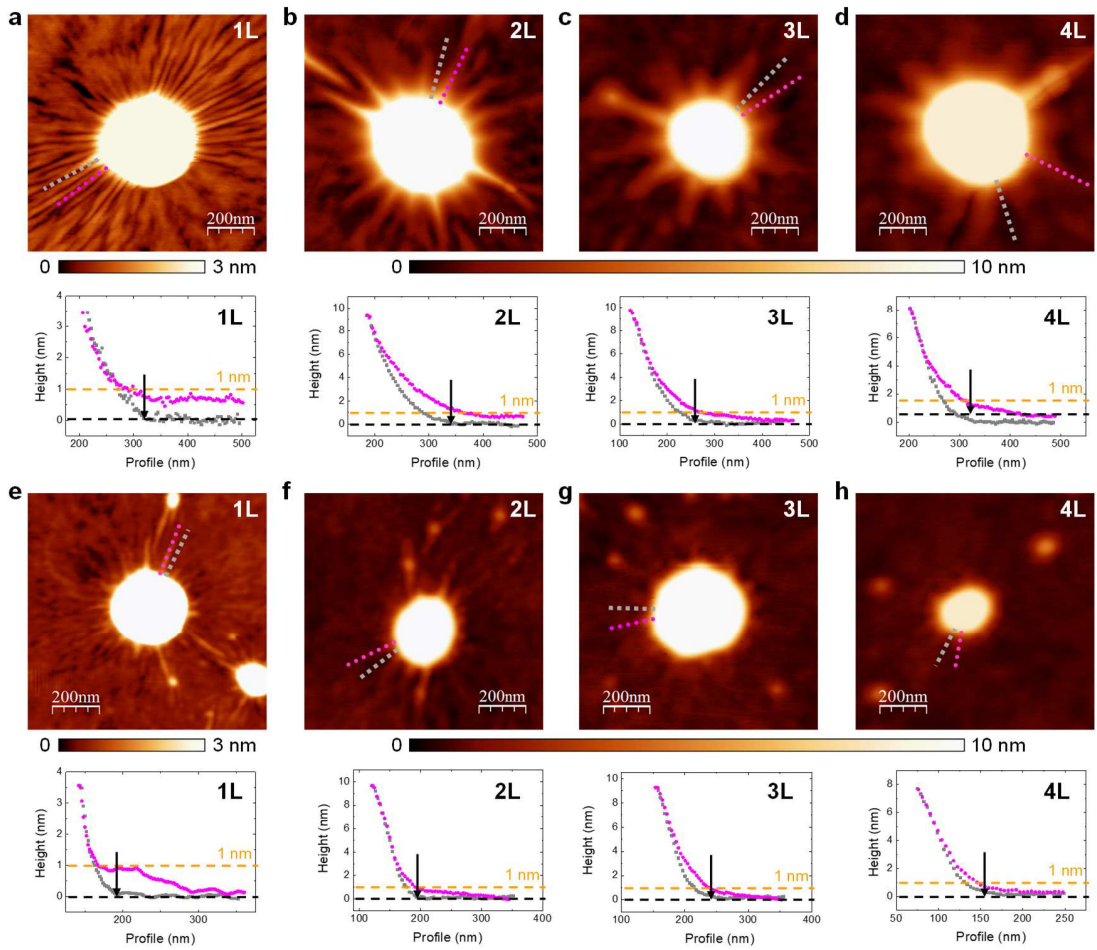

**Fig. S3. Topographical images (top rows) and corresponding representative profiles (bottom rows) of wrinkles. a-d**, hBN and **e-h**, MoS2 wrinkles in layers of different thicknesses. Height scales: 3 nm for 1L and 10 nm for the rest of the cases. The gray squared-dotted lines correspond to the substrate whereas the pink circled-dotted lines to wrinkles. The arrows in the profiles indicate when the bubble reaches the substrate level in the areas without wrinkles. In all the cases, the height of the wrinkles decays exponentially, such that most of the wrinkled zone is nearly flat in the radial direction and smaller than 1 nm in the majority of its length.

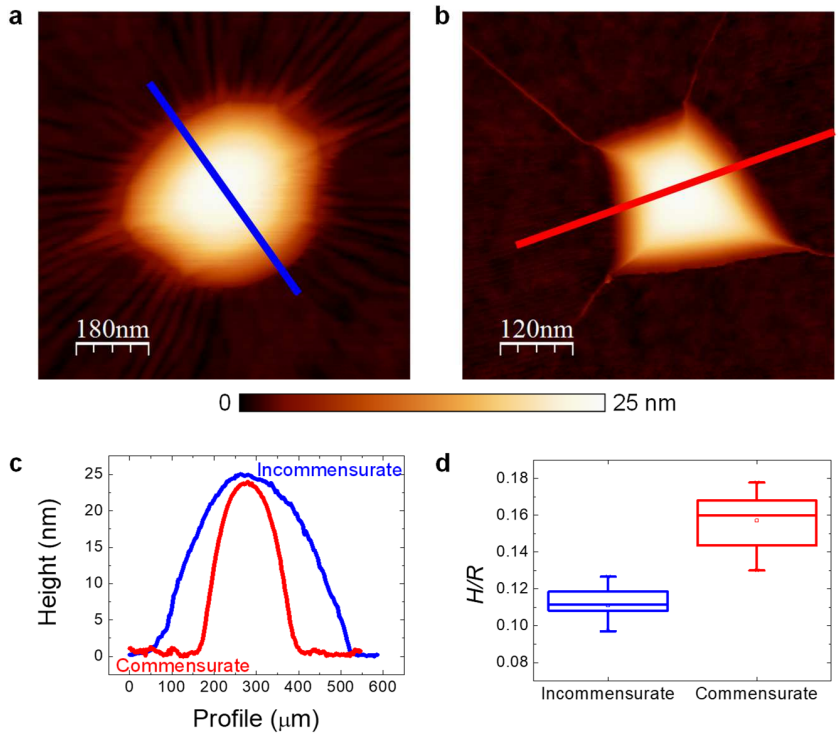

**Fig. S4. Comparative of aspect ratio in bubbles present in incommensurate and commensurate states. a**, Representative bubble in incommensurate and **b**, commensurate states. Height scale: 25 nm. **c**, Profiles along the lines in **a** and **b** showing a significant different aspect ratio for the two bubbles. **d**, Statistical analysis of the aspect ratio for the incommensurate and commensurate states.

Table S1. Elastic constants used for hBN and MoS2.

|  | hBN | MoS2 |
|---|---|---|
| $\lambda$ (eVÅ$^{-2}$) | 3.68 (1) | 2.83 (2) |
| $\mu$ (eVÅ$^{-2}$) | 7.80 (1) | 3.135 (2) |
| $d$ (Å) | 3.34 | 6.5 |
| $B$ (eV) | 0.85 (3) | 10 (4) |

**Substrate-induced stiffness of a membrane on a supporting substrate**

The effective stiffness $K_{eff}(r)$ can originate from multiple contributions (5):

$$K_{eff}(r) = K_{sub} + K_{tens}(r) + K_{curv}(r) \qquad (1)$$

Where $K_{eff}(r)$ ([force]/[length]$^3$, alternatively [energy]/[length]$^4$) acts to suppress the defection (amplitude) of wrinkles from the rest (naturally planar) state of the membrane. Generally, the effective stiffness, $K_{eff}(r)$, comprises three types of restoring (amplitude-suppressing) forces:

i. A substrate-induced stiffness, $K_{sub}$, attributed to an energetic cost, $K_{sub}A^2$, emanates from the normal force exerted on the membrane by a supporting substrate, $K_{sub} = V_{vdW}''(h_0)$. For polymer sheets, such a normal force reflects the energetic cost required to deform the substrate (*e.g.* Young's modulus of a compliant solid substrate (6), or the *g.p.e.* of a liquid bath). In our study, we can consider the supporting substrate as infinitely rigid and hence follow Zhang-Witten (7), assuming that the only normal force exerted by the substrate on the membrane originates from the change in the vdW energy. Namely, $K_{sub} \cdot A = V_{vdW}''(h_0) \cdot A$ is the force required to bring a unit area of the membrane to a distance $h = h_0 + A$ from the substrate (where $h = h_0$ is the minimal-energy distance of a flat membrane from the substrate). Hence, $K_{sub} = V_{vdW}''(h_0)$.
ii. A tension-induced stiffness, $K_{tens}(r)$, is attributed to an energetic cost, $K_{tens}A^2 \propto \sigma_{rr} A'(r)^2$, of radial variation of the wrinkle amplitude, $A(r)$, in the presence of radial tension $\sigma_{rr} \sim \Gamma^{1/2} Y^{1/2}$, *along* wrinkles (8). Since the radial variation of the wrinkle amplitude occurs over a scale $\propto R$, we may estimate, $K_{tens} \sim \Gamma^{1/2} Y^{1/2}/R^2$.
iii. A curvature-induced stiffness, $K_{curv}(r) \sim Y [z_0''(r)]^2$, is attributed to an energetic cost of the strain associated with deforming a curved "envelope" shape, $z_0(r)$, with radial curvature $\sim z_0''(r)$, due to the Gaussian curvature induced by superimposing on it azimuthal undulations (5).

Our experimental observations enable us to rule out the relevance of $K_{curv}(r)$ and $K_{tens}(r)$. First, observation *(c)* in the main text (the non-oscillatory component of the wrinkles decays exponentially, such that most of the wrinkled zone is nearly flat in the radial direction, and highly curved in the azimuthal direction) implies that, while $K_{curv}(r)$ may be significant very close to the bubble, it decays exponentially rapidly, and does not affect the membrane in the vast part of the wrinkled zone. Second, observation *(e)* (for a given type and number of layers of the top membrane, the average wavelength $\ell_0$ does not depend on the size (radius or height) of bubbles) implies that $K_{tens}(r)$ is also negligible, otherwise the average wrinkle wavelength would exhibit a strong dependence on the bubble's radius ($\ell_0 \propto \sqrt{R}$). Finally, observation *(b)* (the wrinkle number $m(r)$ increases with radial distance *r*) indicates a spatially-uniform wavelength, consistently with $K_{eff} \approx K_{sub}$.

A straightforward relation $K_{eff} \sim \Gamma d_{min}^{-2}$ could be used to estimate a characteristic width of the vdW potential well. This expression leads to length scales of the order of 1 nm for the commensurate case, and 10 nm for the incommensurate one. These estimates, which do not seem realistic, will be significantly reduced if one considers, as seems likely, that the very amplitude of the wrinkles and the thermal fluctuations of the membranes at room temperature smooth the vdW potential.

**Bending rigidity of stacks of weakly coupled layers**

The bending rigidity of isotropic, or almost isotropic, elastic slabs is proportional to the in-plane bulk modulus, and it increases as the cube of the thickness of the slabs (9). A stack of layers coupled by the van der Waals interaction has highly anisotropic elastic constants, and a different response to bending.

The coupling between neighboring layers, at long wavelengths, can be described by three force constants, which determine i) the energy required for a change in the interlayer distance, that is, the breathing mode, ii) the energy required for a lateral displacement of one layer with respect the other, the shear mode, and iii) the coupling between a compression within the layer and a change in the interlayer distance, which determines the out of plane Poisson ratio.

The coupling between in plane deformations and bending in a slab arises from the shear deformation which arises when the slab is bent. Hence, it vanishes when the layers can freely slide one over the other. Neglecting the third coupling mentioned above, the elastic energy of a slab of $N$ layers can be written as (10):

$$E = \sum_i \int d^2\vec{r} \left\{ \frac{\lambda}{2} \left( \partial_x u_x^i + \partial_y u_y^i \right)^2 + \mu \left[ (\partial_x u_x^i)^2 + \left( \partial_y u_y^i \right)^2 + \frac{1}{2} \left( \partial_x u_y^i + \partial_y u_x^i \right)^2 \right] \right\} +$$
$$\frac{\Gamma_b}{2} \sum_i \int d^2\vec{r}_i \left( \frac{u_z^i - u_z^{i+1}}{d} \right)^2 + \frac{\Gamma_s}{2} \sum_i \int d^2\vec{r}_i \left[ \left( \frac{-u_x^i + u_x^{i+1}}{d} + \frac{\partial_x u_z^i + \partial_x u_z^{i+1}}{2} \right)^2 + \left( \frac{u_y^i - u_y^{i+1}}{d} + \frac{\partial_y u_z^i + \partial_y u_z^{i+1}}{2} \right)^2 \right] +$$
$$+ \frac{B}{2} \int d^2\vec{r} \left[ (\partial_{xx} u_z^i)^2 + (\partial_{yy} u_z^i)^2 \right] \quad (2)$$

Where $\lambda$ and $\mu$ are in plane elastic constants, $\Gamma_b$ and $\Gamma_s$ describes the interlayer breathing and shear stiffness, $B$ is the bending rigidity of each layer, and $d$ is the distance between layers.

The bending rigidity of a multilayer with N layers can be understood by analyzing the normal modes defined using the energy in eq. (2).

We focus on the low energy modes, described by smooth variations of the displacements between adjacent layers. A simple ansatz which is consistent with the continuum limit of isotropic slabs, which describes normal modes of eq. (2) is:

$$u_z^i(x) = h \sin(Gx), \quad i = 1, \cdots, N$$
$$u_x^i(x) = \left( i - \frac{N+1}{2} \right) u_0 \cos(Gx), \; i = 1, \cdots, N \quad (3)$$

with $G = G_x$, and $h$ and $u_0$ are the amplitudes of out of plane and in plane displacements. The deformation described in eq. (3) does not require a change in the distance between layers, so that it does not depend on the out of plane elastic stiffness, $\Gamma_b$. The insertion of eq. (3) into eq. (2) leads to two modes for each value of $G$, depending on the relative sign between $h$ and $u_0$. We consider the mode of lowest energy, $\omega_{fl}(G)$, whose dispersion is:

$$M\omega_{fl}^2(G)/(A/2) = \frac{1}{2}\left[\frac{N^3-N}{12}(\lambda + 2\mu)G^2 + NBG^4 + \frac{(N-1)\Gamma_s}{d^2} + (N-1)\Gamma_s G^2\right] -$$
$$- \sqrt{\frac{1}{4}\left[\frac{N^3-N}{12}(\lambda + 2\mu)G^2 - NBG^4 + \frac{(N-1)\Gamma_s}{d^2} - (N-1)\Gamma_s G^2\right]^2 + (N-1)^2 \frac{\Gamma_s^2 G^2}{d^2}} \quad (4)$$

Where $A$ is the area of the unit cell, $M = 2 \times N \times M_C$ is the mass of the unit cell, and $M_C$ is the mass of the carbon atom. Expanding eq. (4), we obtain:

$$M\omega_{fl}^2(G) \approx \begin{cases} \frac{N^3-N}{12}(\lambda + 2\mu)G^4 d^2, & G^2 \ll \frac{12(N-1)\Gamma_s}{(N^3-N)d^2(\lambda+2\mu)} \\ N \times BG^4, & G^2 \gg \frac{12(N-1)\Gamma_s}{(N^3-N)d^2(\lambda+2\mu)} \end{cases} \quad (5)$$

Which allows us to define an effective bending rigidity:

$$B_{eff}(G) \approx \begin{cases} \frac{N^3-N}{12}(\lambda + 2\mu)d^2, & G^2 \ll \frac{12(N-1)\Gamma_s}{(N^3-N)d^2(\lambda+2\mu)} \\ N \times B, & G^2 \gg \frac{12(N-1)\Gamma_s}{(N^3-N)d^2(\lambda+2\mu)} \end{cases} \quad (6)$$

where the wavelength is $\ell = (2\pi)/G$. At long wavelengths, the effective bending rigidity scales as the cube of the number of layers, and it is determined by the in plane bulk modulus, in agreement with the general theory of elastic slabs (9). At short wavelengths, the bending rigidity scales like the number of layers, and it is proportional to the rigidity of a single layer. The crossover is determined by the shear stiffness.

We can use the paradigmatic case of graphene to gain some insight on the values of this crossover length. The elastic constants of single layer graphene are:

$$\lambda \approx 2 \text{ eV Å}^{-2}$$

$$\mu \approx 10 \text{ eV Å}^{-2}$$

$$B \approx 1 \text{ eV} \quad (7)$$

And the interlayer distance is $d \approx 3.5$ Å. The parameter $\Gamma_s$ determines the shear mode of graphite and multilayered samples,

$$\omega_{sh}^2 = \frac{\Gamma_s A}{4 M_C d^2} \quad (8)$$

where $A = (3\sqrt{3}a^2)/2$ is the area of the unit cell of graphene, and $a \approx 1.4$ Å is the distance between neighboring carbon atoms.

The frequency of the shear mode in bilayer graphene has been calculated and measured in Ref. (11), and a related calculation for multiwalled nanotubes can be found in Ref. (12). The Raman frequency of the interlayer shear mode in multilayered samples obtained in Ref. (11) depends on the number of layers, $N$. It ranges between $\omega_{fl} \approx c \times 31$ cm$^{-1}$ for a bilayer and $\omega_{fl} \approx c \times 44$ cm$^{-1}$ for graphite, where $c$ is the velocity of light. The dependence of the shear frequency on the number of layers implies the existence of couplings between layers that are not nearest neighbors, not included in eq. (2). Neglecting this effect, we obtain the value of $\Gamma_s$, and a crossover length, $\ell^*$ between the two regimes described in eq. (6):

$$\Gamma_s \approx 17 \text{ meV Å}^{-2}$$

$$\ell^*(N) = \sqrt{\frac{4\pi^2(N^3-N)d^2(\lambda+2\mu)}{12(N-1)\Gamma_s}} \qquad (9)$$

This length increases as the bulk modulus of a monolayer increases, and it decreases as the interlayer shear stiffness increases. For thick multilayers, we find $\lim_{N\to\infty} \ell^*(N) \approx 240 \times N$ nm, which is more than two orders of magnitude larger than the width of the stack, $W = (N-1) \times d$. For a graphene bilayer, the length is $\ell^* \approx 55$ nm.

In very rigid monolayers, the sliding of the monolayers is favored, and the bending rigidity of the stack is determined by the bending rigidity of each layer. On the other hand, in isotropic systems $\Gamma_s \sim \lambda + 2\mu$, and $\ell^* \sim N \times d$. The bending rigidity is independent of the bending rigidity of the individual layers for length scales larger than the width of the slab.

It is interesting to note that in twisted bilayers with a small twist angle most atoms in different layers are not in registry, and the interlayer shear is very small or it even vanishes, when the resulting Moiré pattern is incommensurate with the atomic lattices (12). Then, the bending rigidity of a small angle twisted graphene bilayer is always twice the bending rigidity of each layer.